# ENCRYPTION OF DATA USING ELLIPTIC CURVE OVER FINITE FIELDS

D. Sravana Kumar[1]    CH. Suneetha[2]    A. ChandrasekhAR[3]

[1]Reader in Physics, SVLNS Government College, Bheemunipatnam,
Visakhapatnam Dt., India
skdharanikota@gmail.com

[2]Assistant Professor in Engineering Mathematics, GITAM University, Visakhapatnam,
India,
gurukripachs@gmail.com

[3]Professor in Engineering Mathematics, GITAM University, Visakhapatnam, India
acs@gitam.edu

*ABSTRACT*

*Cryptography is the study of techniques for ensuring the secrecy and authentication of the information. Public –key encryption schemes are secure only if the authenticity of the public-key is assured. Elliptic curve arithmetic can be used to develop a variety of elliptic curve cryptographic (ECC) schemes including key exchange, encryption and digital signature. The principal attraction of elliptic curve cryptography compared to RSA is that it offers equal security for a smaller key-size, thereby reducing the processing overhead. In the present paper we propose a new encryption algorithm using Elliptic Curve over finite fields.*

## 1. INTRODUCTION

The study of elliptic curves by algebraists, algebraic geometers and number theorists dates back to the middle of the nineteenth century. Elliptic Curve Cryptography (ECC) was discovered in 1985 by Neil Koblitz and Victor Miller [7.12]. Elliptic Curve Cryptographic (ECC) schemes are public-key mechanisms that provide the same functionality as RSA schemes. However, their security is based on the hardness of a different problem, namely the Elliptic Curve Discrete Logarithmic Problem (ECDLP). Most of the products and standards that use public-key cryptography for encryption and digital signatures use RSA schemes. The competing system to RSA [8] is elliptic curve cryptography. The principal attraction of elliptic curve cryptography compared to RSA is that it offers equal security for a smaller key-size. An elliptic curve E over a field R of real numbers is defined by an equation

$$E: y^2 + a_1xy + a_3y = x^3 + a_2x^2 + a_4x + a_6 \ldots\ldots\ldots\ldots\ldots\ldots..(1)$$

Here $a_1, a_2, a_3, a_4, a_6$ are real numbers belong to R, x and y take on values in the real numbers. If L is an extension field of real numbers, then the set of L-rational points on the elliptic curve E is
$E(L) = \{(x, y) \in LXL : y^2 + a_1xy + a_3y - x^3 - a_2x^2 - a_4x - a_6 = 0\} \cup \{\infty\}$ where $\infty$ the point is at infinity. Equation (1) is called Weierstrass equation. Here the elliptic curve E is defined over the field of integers K, because $a_1, a_2, a_3, a_4, a_6$ are integers. If E is defined over the field of integers K, then E is also defined over any extension field of K. The condition ≠ 0 ensures that the elliptic curve is "smooth". i.e., there are no points at which the curve has two or more distinct tangent lines. The point ∞ is the only point on the line at infinity that satisfies the projective form of the Weierstrass equation [1,9,14]. In the present paper for the purpose of the encryption and decryption using elliptic curves it is sufficient to consider the equation of the form $y^2 = x^3 + ax + b$.. For the given values of a and b the plot consists of positive and negative values of y for each value of x. Thus this curve is symmetric about the x-axis.

**1.1 Group laws of E(K):-** Let E be an elliptic curve defined over the field of integers K. There is a chord-and-tangent rule for adding two points in E(K), to give the third point. Together with this addition operation, the set of points of $E_p(a,b)$ forms an abelian group with ∞ , the point at infinity as identity elements.

**1.2 Geometric rules of Addition:-** Let $P(x_1,y_1)$ and $Q(x_2,y_2)$ be two points on the elliptic curve E. The sum R is defined as: First draw a line through P and Q, this line intersects the elliptic curve at a third point. Then the reflection of this point of intersection about x-axis is R which is the sum of the points P and Q. The same geometric interpretation also applies to two points P and –P, with the same x-coordinate. The points are joined by a vertical line, which can be viewed as also intersecting the curve at the infinity point. We therefore have P + (-P) =∞, the identity element which is the point at infinity.

**1.3 Doubling the point on the elliptic curve:-**
First draw the tangent line to the elliptic curve at P which intersects the curve at a point. Then the reflection of this point about x-axis is R. As an example the addition of two points and doubling of a point are shown in the following figures 1 and 2 for the elliptic curve $y^2 = x^3-x$.

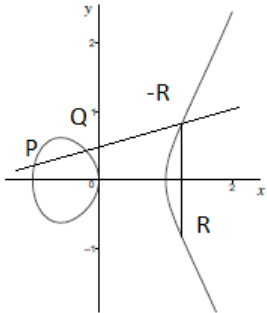 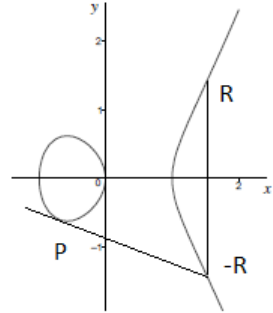

Figure 1. Geometric addition        Figure 2. Geometric doubling

**1.4 Identity:-** P + ∞ = ∞+P = P for all E(K), where ∞ is the point at infinity.

**1.5 Negatives**:- Let P(x,y) ∈ E(K) then (x,y) + (x,-y) = ∞. where (x,-y) is the negative of P denoted by –P.

**1.6 Point addition**:- Let $P(x_1,y_1)$, $Q(x_2,y_2)$ ∈ E(K) where P ≠ Q. Then P + Q = $(x_3,y_3)$ where $x_3 = \left(\dfrac{y_2 - y_1}{x_2 - x_1}\right)^2 - x_1 - x_2$ and $y_3 = \left(\dfrac{y_2 - y_1}{x_2 - x_1}\right)(x_1 - x_3) - y_1$

**1.7 Points Doubling:-** Let $P(x_1, y_1) \in E_K(a,b)$ where $P \neq -P$ then

$$2P = (x_3, y_3) \text{ where } x_3 = \left(\frac{3x_1^2 + a}{2y_1}\right)^2 - 2x_1 \text{ and } y_3 = \left(\frac{3x_1^2 + a}{2y_1}\right)(x_1 - x_3) - y_1$$

**1.8 Point Multiplication:-** Let P be any point on the elliptic curve(K). Then the operation multiplication of the point P is defined as repeated addition. kP = P + P + .........ktimes

**1.9 Elliptic Curve Cryptography**:- Elliptic Curve Cryptography (ECC) [2, 3,4,5,] makes use of the elliptic curve in which the variables and coefficients are all restricted to elements of the finites fields [10]. Two families of elliptic curves are used in cryptographic applications: Prime curves [7,15]over $Z_p$ and binary curves $GF(2^m)$. For a prime curve over $Z_p$, we use a cubic equation in which the variables and the coefficients all take on values in the set of integers from 0 through p-1 and the calculations are performed with respect to modulo p.

**2. Related Work**:- Elliptic curve cryptography has been thoroughly researched for the last twenty years. The actual application of elliptic curve cryptography and the practical implementation of cryptosystem primitives in the real world constitute interdisciplinary research in computer science as well as in electrical engineering. Elliptic Curve Cryptography provides an excellent solution not only for the data encryption but also for the secure key transport between two communicating parties [ 16 ], and authentic session key establishment protocols [6,11,13].
$E_p$ (a,b).

**3. Proposed Method**:- If two communicating parties Alice and Bob want to communicate the messages then they agree upon to use an elliptic curve $E_p$ (a,b) where p is a prime number and a random point C on the elliptic curve. Alice selects a large random number α which is less than the order of $E_p$ (a,b) and a point A on the elliptic curve. She computes $A_1$ = α (C +A) and $A_2$ = α A. She keeps the random number α and the point A as her private keys and publishes $A_1$ and $A_2$ as her general public keys. Similarly Bob selects a large random number β and a point B on the elliptic curve. He computes
$B_1$ = β (C+B) and $B_2$ = β B. He keeps the random number β and the point B as his private keys and publishes $B_1$ and $B_2$ as his general public keys. After publishing the public keys, the communicating parties again calculate the following quantities and publish them as their specific public keys of each other.
Alice calculates $A_B$ = α $B_2$ and publishes it as her specific public key for Bob
Bob calculates $B_A$ = β $A_2$ and publishes it as his specific public key for Alice

| | |
|---|---|
| Alice's private key 1 | = α, a large random number less than the order of the generator |
| Alice's private key 2 | = a point A on the elliptic curve $E_p$ (a,b) |
| Alice's general public key 1 | = a point $A_1$ on the elliptic curve $E_p$ (a,b) |
| Alice's general public key 2 | = a point $A_2$ on the elliptic curve $E_p$ (a,b) |
| Alice's specific public key for Bob | = a point $A_B$ on the elliptic curve $E_p$ (a,b) |
| Bob's private key 1 | = β, a large random number less than the order of the generator |
| Bob's private key 2 | = B, a point on the elliptic curve $E_p$ (a,b) |

International Journal of Distributed and Parallel Systems (IJDPS) Vol.3, No.1, January 2012Bob's general public key 1 = $B_1$, a point on the elliptic curve $E_p(a,b)$
Bob's general public key 2 = $B_2$, a point on the elliptic curve $E_p(a,b)$
Bob's specific public key for Alice = $B_A$, a point on the elliptic curve $E_p(a,b)$

**3.1 Encryption:**- If Bob wants to communicate the message M then all the characters of the message are coded to the points on the elliptic curve using the code table which is agreed upon by the communicating parties Alice and Bob. Then each message point is encrypted to a pair of cipher points $E_1, E_2$. He uses a random number γ which is different for the encryption of different message points.

$E_1 = γ\ C$
$E_2 = M + (β + γ)\ A_1 - γ\ A_2 + A_B$

After encrypting all the characters of the message Bob converts the pair of points of each message point into the text characters using the code table. Then he communicates the cipher text to Alice in public channel.

**3.2 Decryption**:- After receiving the cipher text, Alice converts the cipher text into the points on the elliptic curve and recognizes the points $E_1$ and $E_2$ of each character. Then she decrypts the message as follows.

$M = E_2 - (α\ E_1 + α\ B_1 + B_A)$

**3.3 Decryption works out properly**:- $(β + γ)\ A_1 - γ\ A_2 + A_B = γ\ (A_1 - A_2) + β\ A_1 + A_B$
$= γ\ α\ C + β\ α\ C + β\ α\ A + β\ α\ B$
$= γ\ α\ C + β\ α\ (A+B+C)$
$αE_1 + α\ B_1 + B_A = α\ γ\ C + α\ β\ C + α\ β\ B + α\ β\ A$
$= γ\ α\ C + β\ α\ (A+B+C)$
Therefore, $(β + γ)\ A_1 - γ\ A_2 + A_B = α\ E_1 + α\ B_1 + B_A$
$E_2 - (α\ E_1 + α\ B_1 + B_A = [M + (β + γ)\ A_1 - γ\ A_2 + A_3] - [α\ E_1 + α\ B_1 + B_A]$
$= M + [γ\ α\ C + β\ α\ (A+B+C)] - [γ\ α\ C + β\ α\ (A+B+C)]$
$= M$

In this method a group of communicating parties A, B, C, D……. can communicate with one another securely, non- repudiatively in an authentic manner. Here each communicating party say X publishes two general public keys $X_1, X_2$. X also publishes a specific public key $X_Y$ to be used by the communicating party Y for communication with Y. When Y wants to communicate with X, Y uses the general public keys of X $(X_1, X_2)$, the specific public key published by X for Y $(X_Y)$ and Y's secret key y. To decrypt the message X uses Y's general public keys $(Y_1, Y_2)$, the specific public key published by Y for X $(Y_X)$ and X's secret key x. Here X creates specific public key $X_Y$ for Y using Y's public keys and X's secret key. So, this method of encryption using elliptic curves over finite fields is highly suitable for communication between groups of corporate/government institutions.

**4. Example:**- Consider an elliptic curve whose equation is $y^2 = x^3 + 2x + 9$. The graph of the function is shown in figure 3.

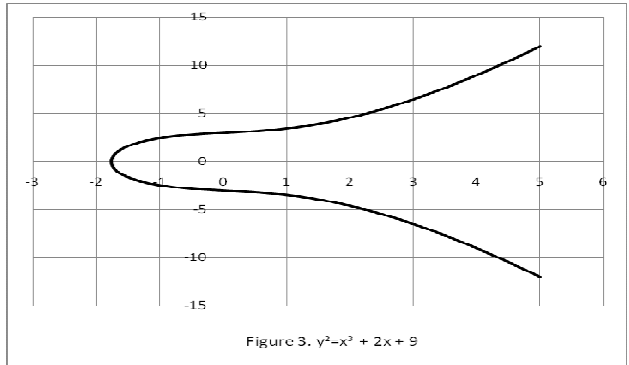

In the above graph the right lines can be drawn in xy-plane such that 1) there is no intersection between the right line and elliptic curve 2) the line intersects the elliptic curve at one point or two points or three points.

Now consider an elliptic curve ($y^2 = x^3 + 2x + 9$)$_{mod37}$ , $E_{37}$ (2,9). The points on the elliptic curve
 $E_{37}$ (2,9) are
{ ∞,(5,25), (1,30), (21,32), (7,25), (25,12), (4,28), (0,34), (16,17), (15,26), (27,32), (9,4),(2,24), (26,5), (33,14), (11,17), (31,22), (13,30), (35,21), (23,7), (10,17), (29,6), (29,31), (10,20), (23,30), (35,16),(13,7), (31,15), (11,20), (33,23), (26,32),(2,13), (9,33), (27,5), (15,11), (16,20), (0,3), (4,9), (25,25), (7,12), (21,5), (1,7), (5,12), }
The graph of the function is shown in Figure 4.

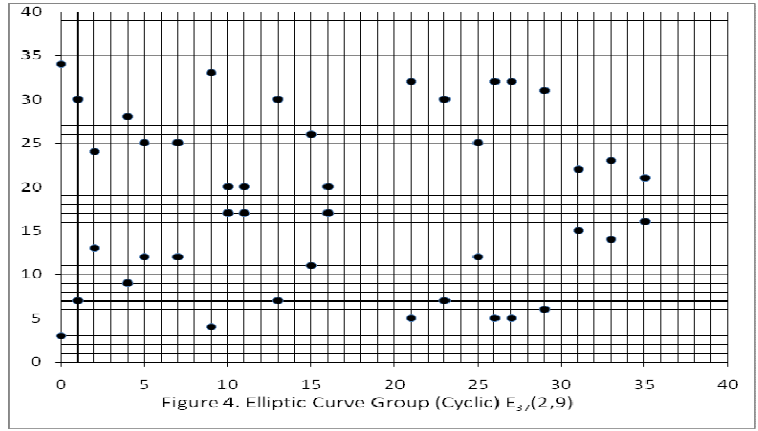

Let C = (9,4). Alice selects a random number α = 5, any point A = (10,20) on the elliptic curve. She computes
 $A_1$ = α (C+A) = 5[(9,4) + (10,20)] = (1,7)
 $A_2$ = α A = (33,23).
She keeps the random number α = 5 and the point A on the elliptic curve as her secret keys and publishes $A_1$ and $A_2$ as her public keys.
Bob selects β = 7, B = (11,20) on the elliptic curve. He computes
 $B_1$ = β (C+B) = (11,17)
 $B_2$ = β B = (23,30).
He keeps the random number β = 7 and the point B on the elliptic curve as his secret keys and publishes $B_1$ and $B_2$ as his public keys.

Alice calculates $A_B = \alpha B_2 = (15,11)$ and Bob calculates $B_A = \beta A_2 = (2,13)$. Alice publishes $A_B$ as the specific public key for Bob and Bob publishes $B_A$ as specific public key for Alice.

**4.1 Encryption**: If Bob wants to communicate the message 'attack' to Alice, Bob converts all the text characters of the message into the points on the elliptic curves using the agreed upon code table.

| * | a | b | c | d | e | f | g | h |
|---|---|---|---|---|---|---|---|---|
| ∞ | (5,25) | (1,30) | (21,32) | (7,25) | (25,12) | (4,28) | (0,34) | (16,17) |
| I | j | k | l | m | n | o | p | q |
| (15,26) | (27,32) | (9,4) | (2,24) | (26,5) | (33,14) | (11,17) | (31,22) | (13,30) |
| r | s | t | u | v | w | x | y | z |
| (35,21) | (23,7) | (10,17) | (29,6) | (29,31) | (10,20) | (23,30) | (35,16) | (13,7) |

| 1 | 2 | 3 | 4 | 5 | 6 | 7 | 8 | 9 | 0 |
|---|---|---|---|---|---|---|---|---|---|
| (31,15) | (11,20) | (33,23) | (26,32) | (2,13) | (9,33) | (27,5) | (15,11) | (16,20) | (0,3) |
| # | @ | ! | & | $ | % | | | | |
| (4,9) | (25,25) | (7,12) | (21,5) | (1,7) | (5,12) | | | | |

1). In the message 'attack' the first character 'a' corresponds to the point (5,25) using the code table. Bob selects a random number $\gamma = 8$ for encrypting the character 'a'. Then the point (5,25) is encrypted as

$E_1 = \gamma C = (1,30)$ which corresponds to the character 'b' in the conversion table.

$E_2 = M + (\beta + \gamma) A_1 - \gamma A_2 + A_B = (2,13)$ which corresponds to '5' in the code table. So, the character 'a' in the plain text is encrypted to two characters {b,5} in the cipher text.

2) 't' is a point (10,17) in the code table. Let $\gamma = 12$

$E_1 = (21,32)$ which corresponds to 'c' in the code table.

$E_2 = M + (\beta + \gamma) A_1 - \gamma A_2 + A_B = (2,24)$ which corresponds to 'l' in the code table. So, 't' is encrypted as {c,l}.

3) 't' is a point (10,17) in the code table. Let $\gamma = 19$

$E_1 = (4,9)$ which corresponds to '#' in the code table.

$E_2 = M + (\beta + \gamma) A_1 - \gamma A_2 + A_B = (27,32)$ which corresponds to 'j' in the code table. So, 't' is encrypted as {#,j}

4) 'a' is a point (5,25) in the code table. Let $\gamma = 2$

$E_1 = (29,31)$ which corresponds to 'v' in the code table.

$E_2 = M + (\beta + \gamma) A_1 - \gamma A_2 + A_B = (1,30)$ which corresponds to 'b' in the code table. So, 'c' is encrypted as {v,b}

5) 'c' is a point (21,32) in the code table. Let $\gamma = 3$
$E_1 = (1,30)$ which corresponds to 'b' in the code table.

$E_2 = M + (\beta + \gamma) A_1 - \gamma A_2 + A_B = (31,22)$ which corresponds to 'p' in the code table. So, 'a' is encrypted as {b,p}.

6) 'k' is a point (9,4) in the code table. Let $\gamma = 23$

$E_1 = (25,25)$ which corresponds to '@' in the code table.

$E_2 = M + (\beta + \gamma) A_1 - \gamma A_2 + A_B = (4,28)$ which corresponds to 'f' in the code table. So, 'k' is encrypted as {@,f}

Bob communicates {b,5; c,l; #,j; v,b; b,p; @,f }as the cipher text to Alice in public channel.

**4.2 Decryption**:- Alice after receiving the cipher text {b,5; c,l; #,j; v,b; b,p; @,f } converts the cipher characters into the points (1,30) ,(2,13), (21,32),(2,24) (4,9) ,(27,32) (29,31) (1,30) (1,30) (31,22) (25,25) (4,28). She decrypts the message taking two points at a time as the points E1 and E2.

1) $M = E_2 - (\alpha E_1 + \alpha B_1 + B_A) = (5,25)$ which corresponds to the character 'a' in the code table.
2) $M = E_2 - (\alpha E_1 + \alpha B_1 + B_A) = (10,17)$ which corresponds t the character 't' in the code table.
3) $M = E_2 - (\alpha E_1 + \alpha B_1 + B_A) = (10,17)$ which corresponds to the character 't' in the code table.
4) $M = E_2 - (\alpha E_1 + \alpha B_1 + B_A) = (5,25)$ which corresponds to the character 'a' in the code table.
5) $M = E_2 - (\alpha E_1 + \alpha B_1 + B_A) = (21,32)$ which corresponds to the character 'c' in the code table.
6) $M = E_2 - (\alpha E_1 + \alpha B_1 + B_A) = (9,4)$ which corresponds to the character 'k' in the code table. Then 'attack' is the original message.

## 5. Conclusions:-

In the encryption algorithm proposed here the communicating parties agree upon to use an elliptic curve and a point C on the elliptic curve. The security of the Elliptic Curve Cryptography depends on the difficulty of finding the value of k, given kP where k is a large number and P is a random point on the elliptic curve. This is the Elliptic Curve Discrete Logarithmic Problem. The elliptic curve parameters for cryptographic schemes should be carefully chosen in order to resist all known attacks of Elliptic Curve Discrete Logarithmic Problem (ECDLP). The straightforward use of public key encryption provides confidentiality but not the authentication [17]. Each communicating party publishes a specific public key for the communication with a specific communicator. With this the receiver is assured that the cipher was constructed by the sender only because the sender uses receiver's general public keys, receiver's specific public key published for the sender alone and sender's private key for constructing the cipher. This ensures that sender has "digitally signed" the message by using the specific public key published for him alone by the receiver. Hence, the cipher has achieved the qualities confidentiality, authentication and non-repudiation. Moreover, each message point is encrypted as a pair of points on the elliptic curve. Here a random number γ is used in the encryption of each message point and γ is different for encryption of different message points. That is why the same characters in the message space are encrypted to different characters in the cipher space. The difference between characters of the plain text is not the same as difference between the characters of the cipher text. Due to this the linear cryptanalysis is highly difficult. In addition to this each character of the message is coded to the point on the elliptic curve using the code table which is agreed upon by the communicating parties and each message point is encrypted to a pair of points on the elliptic curve. Hence, the method of encryption proposed here provides sufficient security against cryptanalysis at relatively low computational overhead.

## 6. References:-

Dr. D. Sravana Kumar is a senior faculty in Physics in Government College, Visakhapatnam. He obtained his doctorate in Ultrasonics. His research interest includes Ultrasonics, Molecular Interactions and Cryptography. He has published 14 research papers in various international journals those published by Springer, Elsevier, etc. He is an ex-scientist in the Department of Atomic Energy, Government of India. He is very much interested in interdisciplinary research.

CH. Suneetha is Assistant Professor in Engineering Mathematics in GITAM University, Visakhapatnam. She obtained her master's degree in applied mathematics, M.Phil. in algebra. At present she is pursuing her Ph.D under the guidance of Dr. A. Chandrasekhar.Her research interests include Cryptography, Linear Transformations and Algebraic Curves.

Dr. A. Chandrasekhar is Professor and Head of the Department of Engineering Mathematics in GITAM University, Visakhapatnam. He obtained his doctorate in Cryptography. His research interest includes Cryptography and Fixed Point Theory. He has published 20 research papers in various reputed national and international journals including IEEE.